\documentclass[aps,prd,twocolumn,superscriptaddress,nofootinbib]{revtex4-2}
\usepackage{amsmath}
\usepackage{amsfonts,amssymb}
\usepackage{mathrsfs}
\usepackage{graphicx,graphics}
\usepackage[hidelinks]{hyperref}
\usepackage{bm} 

\newcommand{\eq}[1]{Eq.~(\ref{#1})}
\newcommand{\eqs}[1]{Eqs.~(\ref{#1})}
\newcommand{\eb}{\bar{e}}

\usepackage{xcolor}

\DeclareMathOperator{\arctanh}{arctanh}

\begin{document}

\title{Chaotic dynamics of a spinless axisymmetric extended body around a Schwarzschild black hole}

\author{Ricardo A. Mosna}
\email{mosna@unicamp.br}
\affiliation{Departamento de Matem\'atica Aplicada, Universidade Estadual de Campinas, 13083-859,  Campinas,  S\~ao Paulo,  Brazil}   
\author{Fernanda F. Rodrigues}
\email{fariafis@ifi.unicamp.br}
\affiliation{Instituto de Física ``Gleb Wataghin", Universidade Estadual de Campinas, 13083-859, Campinas,  S\~ao Paulo, Brazil}
\author{Ronaldo S. S. Vieira}
\email{ronaldo.vieira@ufabc.edu.br}
\affiliation{Centro de Ci\^encias Naturais e Humanas, Universidade Federal do ABC, 09210-580 Santo Andr\'e, SP, Brazil}
\date{\today}

\begin{abstract}
We investigate the long-term orbital dynamics of spinless extended bodies in Schwarzschild geometry, and show that periodic deviations from spherical symmetry in the shape of a test body may trigger the onset of chaos. We do this by applying Dixon's formalism at quadrupolar order to a nearly spherical body whose shape oscillates between a prolate and an oblate spheroid. The late-time chaotic behavior 
is then verified by applying Melnikov's method.
\end{abstract}

\maketitle

\section{Introduction }
\label{Introduction} 

A realistic description of the dynamics of extended bodies is fundamental to the understanding of several physical phenomena. For instance, in Newtonian gravity, an extended body controlling its internal structure can actively modify its orbit \cite{reloc, react, tether, shape}, as well as stabilize it \cite{stab}. The qualitative long-term behavior of a system can also be deeply affected as finite-size corrections are taken into account. It was recently shown that this may even trigger the onset of chaos in an otherwise integrable dynamics of a test particle~\cite{bolinha}. It is to be expected that this fact is all the more pronounced in relativistic physics, partly due to its intrinsic nonlinear nature. 

In this paper, we explore how extended-body effects in general relativity may affect the late-time dynamics of a body by producing chaotic motion, even in situations where this would not be expected in Newtonian mechanics. We do this by applying Dixon's formalism~\cite{dixonI, dixonII, dixonIII} at the quadrupolar order to analyze  the motion of a spinless extended body in the Schwarzschild background, wherein all the point-particle trajectories (its geodesics) are known and regular. Our key finding is that periodic perturbations to the internal structure of a spinless and nearly spherical body can generate chaos in an otherwise regular translational motion. Bodies of this kind may be seen as prototypes of spacecrafts or as toy models for astronomical objects, such as pulsating stars and artificial satellites with internal motion. 

This work is organized as follows. Section \ref{Dixon} reviews some aspects of Dixon's formalism. Section \ref{Schwar} studies the dynamics of extended bodies up to quadrupolar order and prepares the ground for Sec. \ref{melnikov}, wherein we apply Melnikov's method to show the onset of chaos on the motion of an extended body with small periodic deviations from spherical symmetry. We pay special attention to the case when the body is a nearly spherical ellipsoid whose shape oscillates between a prolate and a oblate spheroid. Section~\ref{sec:conclusion} presents our conclusion and final remarks.

\section{Dixon's formalism for quadrupolar bodies}
\label{Dixon} 

Dixon's formalism is a covariant framework to deal with extended bodies on curved spacetimes. It was proposed by W. G. Dixon in a series of articles \cite{dixonI, dixonII, dixonIII} in the seventies; a modern review can also be found at Ref.~\cite{harte2015motion}. It provides both a conceptually satisfying and a powerful calculation tool to deal with the problem of extended bodies in general relativity by developing a set of covariant equations of motion for their linear momentum $p_\mu$ and spin tensor $S^{\mu\nu}$. This is done by means of a (very carefully chosen) multipolar expansion of the body's energy-momentum tensor.

Consider an extended body described by the energy-momentum tensor $T^{\alpha \beta}$ in a spacetime $\mathcal{M}$ with metric $g_{\alpha\beta}$. We assume that this is a test body, {\it i.e.}, that the backreaction of  $T^{\alpha \beta}$ on the metric  of $\mathcal{M}$ is negligible. Its equations of motion are then given by
\begin{equation}\label{divT=0}
\nabla_{\beta}T^{\alpha \beta}=0.
\end{equation}

In Dixon's formalism, the energy-momentum tensor $T^{\alpha \beta}$ is expressed in terms of a set of multipole moments associated with it. The four partial differential equations in (\ref{divT=0}) are then shown to be equivalent to a set ordinary differential equations (ODEs) for these multipoles. Crucial to this method is a very careful choice of a spacetime foliation near the body by spacelike three-surfaces $\Sigma_s$ along its center-of-mass worldline, $z^\mu(s)$. The latter is implicitly defined by the condition
 \begin{equation}
S^{\mu \nu}(s) p_{\nu}(s)=0, 
\end{equation}
where $s$ is an evolution (timelike) parameter. Each $\Sigma_s$ is given by the three-surface that is orthogonal to $p^{\mu}(s)$ at $z^\mu(s)$ and is generated by all the spacelike geodesics emanating orthogonally to $p^{\mu}(s)$ from $z^\mu(s)$.

With these particular choices (for the center of mass and foliation) associated with the body, \eq{divT=0} determines ODEs for the evolution of the monopole ($p_{\mu}$) and dipole ($S^{\mu\nu}$) moments, while leaving all the remaining  (quadrupole, octupole, etc) moments arbitrary. The latter can be freely chosen as reflecting internal details of the body, such as its constitutive laws or inner mechanisms. This leads to a remarkable simplification of the problem. 

The new equations of motion are then given by ten ODEs  for the evolution of the momentum $p_{\mu}(s)$ and spin $S^{\mu\nu}(s)$ of the body:
\begin{subequations}
\label{eq:p&S}
\begin{align}
\frac{Dp^{\mu}}{ds} &= -\frac{1}{2}R^{\mu}_{\phantom{\mu} \nu \alpha \beta}v^{\nu}S^{\alpha \beta}+F^{\mu}, \label{eq:p} \\ 
\frac{DS^{\mu \nu}}{ds} &=2p^{[\mu}v^{\nu]}+N^{\mu \nu}, \label{eq:S}
\end{align}
\end{subequations}
where $R^{\mu}_{\phantom{\mu} \nu \alpha \beta}$ is the Riemann tensor of $g_{\alpha\beta}$ and $v^\mu=dz^\mu/ds$ is tangent to the center-of-mass worldline. 

The terms $F^{\mu}$ and $N^{\mu \nu}$ can be seen as the force and torque with respect to the center of mass, respectively, that act on the body, and may be obtained from \eq{divT=0} in terms of the multipole expansion discussed above. Their expressions up to quadrupole order are given by
\begin{subequations}
\label{eq:F&N}
\begin{align}
F^{\mu} &=- \frac{1}{6}J^{\alpha \beta \gamma \delta}\nabla^{\mu}R_{\alpha \beta \gamma \delta}, \label{eq:F} \\ 
N^{\mu \nu} &=\frac{4}{3}J^{\alpha \beta \gamma[\mu}R^{\nu]}_{\phantom{ai} \gamma \alpha \beta}, \label{eq:N}
\end{align}
\end{subequations}
where $J^{\alpha \beta \gamma \delta}$ is the quadrupole moment of the body~\cite{dixonI, dixonII}. This tensor conveys all the information about the internal structure of the body in \eqs{eq:F&N} and thus, as long as the usual energy conditions are satisfied, it may be prescribed at will [in the sense that any prescription is always consistent with Eq.~\eqref{divT=0}]. Moreover, the quadrupole tensor may be chosen so that it has the same algebraic properties as the Riemann tensor,
\begin{equation}
J^{\alpha \beta \gamma \delta}=J^{[\alpha \beta] \gamma \delta}=J^{\alpha \beta [\gamma \delta]}, \quad \quad J^{[\alpha \beta \gamma] \delta}=0.
\end{equation}

Symmetries of the spacetime give rise to conserved quantities constructed from the monopole and dipole moments. More precisely, one can show that each Killing vector field $\xi$ on $\mathcal{M}$ gives rise to the quantity
\begin{equation}
\label{eq:cons_qtity}
\mathscr{P}_{\xi}=p_{\mu}\xi^{\mu}+ \frac{1}{2}S^{\mu \nu}\nabla_{\mu}\xi_{\nu}
\end{equation}
that is conserved by the time evolution generated by \eqs{eq:p&S}. This conservation law holds exactly ({\it i.e.}, regardless of the quadrupole approximation) and at every perturbation order~\cite{Ehlers1977DynamicsOE}. 

The evolution parameter $s$ is so far arbitrary. 
Let $u^\mu$ be the unit vector in the direction of $p^{\mu}$, so that $p^\mu= m u^\mu$ with $p_\mu p^\mu=-m^2$. The mass $m$ may vary with time when the body is extended; also, $u^\mu\neq v^\mu$ in general.
Usual choices for the evolution parameter $s$ are the time $\tilde{\tau}$ as measured by an observer momentarily at rest with respect to $p^\mu$, so that $v_\mu u^\mu=-1$~\cite{stab},
and the proper time $\tau$ along the center-of-mass trajectory~\cite{2016PhRvD..94l1502S}, so that $v_\mu v^\mu=-1$.
From now on we choose to take $s=\tau$.

We are concerned here with the motion of an extended body in a Schwarzschild black hole spacetime, with metric
\begin{equation}\label{schwmetric}
ds^2=-f(r) dt^2+f(r)^{-1}dr^2+ r^2(d\theta^2+ \sin^2 \theta \, d\phi^2),
\end{equation}
where $f(r)=1-2M/r$ and $M$ is the mass of the black hole. Since this spacetime is static and spherically symmetric, it possesses four independent Killing vectors,
which give rise to four conserved quantities according to \eq{eq:cons_qtity}.

It will also be useful in what follows to define the tetrad $\{ e_{\hat{t}},e_{\hat{r}},e_{\hat{\theta}},e_{\hat{\phi}} \}$ constructed from the coordinate basis $\{ \partial_\mu \}$ as
\begin{equation}\label{basis}
e_{\hat{t}}= \frac{1}{\sqrt{f(r)}}\partial_t,\ \  e_{\hat{r}}=\sqrt{f(r)}\partial_{r},\ \  e_{\hat{\theta}}=\frac{1}{r}\partial_{\theta},\ \ e_{\hat{\phi}}=\frac{1}{r\sin \theta}\partial_{\phi}.
\end{equation}

\section{\label{Schwar} Dynamics of an axisymmetric body in the Schwarzschild spacetime }

It is a well-established fact that spinning particles in Schwarzschild spacetime may have chaotic behavior if their initial spin is not aligned with the equatorial plane, see for instance Ref. \cite{suzukiMaeda1997PRD}. Here we are interested in another kind of phenomenon and, to emphasize this, we study only spinless bodies in what follows.

We consider an extended test body which controls its shape in such a way that it is always axisymmetric and has reflection symmetry with respect to the equatorial plane. This is implemented by choosing an appropriate prescription for the quadrupole moment of the body, $J^{\alpha \beta \gamma \delta}(\tau)$, as discussed below. Assume that the body starts with $S^{\mu\nu}=0$. Then it will remain spinless at all times by symmetry.  As a result, Eq.~(\ref{eq:p}) simplifies to 
\begin{equation}
\label{eq:psimplificada} 
\frac{Dp_\mu}{d\tau}=F_\mu,
\end{equation}
for $\mu=t$, $r$ and $\phi$ (three equations), since the motion is restricted to a plane (chosen as the equatorial plane $\theta=\pi/2$). Equation~(\ref{eq:S}) is identically satisfied with $S^{\mu\nu}\equiv0$ by symmetry (this, of course, can always be checked later).%
\footnote{This does not mean that the body is torque free, as observed in~\cite{stab}. In fact, the ``electric part'' $N^{0i}$ of the torque still contributes to the hidden momentum of the body.
}

We note that Eq.~(\ref{eq:psimplificada}) is linear in $\dot{t}(\tau)$, $\dot{r}(\tau)$ and $\dot{\phi}(\tau)$ and so can be easily rewritten in the form
\begin{subequations}
\label{eq:U=something}
\begin{align}
\dot{t}(\tau) &= h_1(r,p_r,\dot{p}_r,\tau), \label{eq:Ut=something}\\ 
\dot{r}(\tau)  &= h_2(r,p_r,\dot{p}_r,\tau), \label{eq:Ur=something}\\ 
\dot{\phi}(\tau)  &= h_3(r,p_r,\dot{p}_r,\tau), \label{eq:Uphi=something}
\end{align}
\end{subequations}
where the dot represents $d/d\tau$. 
These equations depend explicitly on $\tau$ only via the quadrupole tensor $J^{\alpha \beta \gamma \delta}(\tau)$.
Note that $p_t$ and $p_\phi$ are constants of motion due to \eq{eq:cons_qtity} (with $\xi$ equal to $\partial_t$ and $\partial_\phi$, respectively) and therefore only appear in the above equations parametrically. Moreover, also by symmetry, none of the functions $h_i$ depend on $t$ and $\phi$ (this too can of course be explicitly checked). Equations (\ref{eq:U=something}), together with the constraint $v_\mu v^\mu=-1$, provide four ODEs to determine the unknown functions $t(\tau)$, $r(\tau)$, $\phi(\tau)$ and $p_r(\tau)$.

By substituting \eqs{eq:U=something} into the constraint $v_\mu v^\mu=-1$ we get an equation of the form $\dot{p}_r=h_4(r,p_r,\tau)$. This can be substituted in \eq{eq:Ur=something} to get an equation of the form $\dot{r}=h_5(r,p_r,\tau)$. As a result, we end up with a non-autonomous dynamical system of effectively 1 degree of freedom, defined by
\begin{subequations}
\label{eq:sistdyn}
\begin{align}
\dot{p}_r &= h_4(r,p_r,\tau), \\
\dot{r} &= h_5(r,p_r,\tau).
\end{align}
\end{subequations}

\subsection{Symmetries of the body}

What complicates matters in the procedure just outlined is that, to implement the conditions that the body is axisymmetric and has reflection symmetry with respect to the equatorial plane, we must first go to a moving frame $\{\eb_a\}$ that moves along with the center of the mass of the body. It is in this frame that the expression of the quadrupole moment has axial and reflection symmetry. 

Following Dixon, we take this moving frame by choosing $\eb_0$ as the unit vector in the direction of $p$ ({\it i.e.} $\eb_0=u$) and $\{\eb_1, \eb_2, \eb_3\}$ a nonrotating orthonormal triad in the rest space of $\eb_0$. The choice of $\eb_0$ in the direction of $p$ instead of $v$ is due to the fact that it is $p$ that defines the surfaces of integration $\Sigma_\tau$ of the multipoles ($v$ and $u$ may be thought of as the kinematical an dynamical velocities, respectively) . Also, by ``nonrotating'' we mean that the vectors $\eb_a$ are M-transported along the center-of-mass curve~\cite{dixonI}, so that they are four-vectors $A^\mu$ which satisfy the equation of motion%
\footnote{
Physically speaking, an arbitrary body subjected to a zero torque ($N^{\mu\nu}=0$) would have its spacelike spin vector $S_\mu=\frac{1}{2}\sqrt{-g}\epsilon_{\mu \nu \gamma \delta}u^{\nu}S^{\gamma \delta}$ satisfying the M-transport equation. In this way, an M-transported $\{\eb_{a}\}$ provides a physical standard of (absence of) rotation.
}
\[
\frac{DA^\kappa}{d\tau}-\left( u^\kappa \frac{Du_\lambda}{d\tau} - \frac{Du^\kappa}{d\tau} u_\lambda \right) A^\lambda=0.
\]

If the worldline is restricted to lie on the equatorial plane of Schwarzschild spacetime, one may take from start $\eb_0=u$, $\eb_3=-e_{\hat{\theta}}$, so that only $\eb_1$ and $\eb_2$ need to be determined. The assumption that the body is axisymmetric and has reflection symmetry with respect to the equatorial plane then implies that only six independent components of the quadrupole tensor may be nonzero in this (M-transported) frame:  $\bar{J}_{0101}=\bar{J}_{0202}$, $\bar{J}_{0303}$, $\bar{J}_{2323}=\bar{J}_{1313}$, and $\bar{J}_{1212}$ (terms like $\bar{J}_{0123}$ would be allowed by rotation symmetry but not by the reflection symmetry).

From a more practical point of view, it is easier to construct another moving frame $\{ e_a\}$ along the center-of-mass worldline $z^\mu(\tau)$, one that is not M-transported but also satisfies $e_0=u$ and $e_3=-e_{\hat{\theta}}$. Any frame of this kind can only differ from $\{\eb_a\}$ by a rotation in the plane of $\eb_1$ and $\eb_2$, so that $\eb_1=\cos(\beta) e_1+\sin(\beta) e_2$, $\eb_2=-\sin(\beta) e_1+\cos(\beta) e_2$, with $\beta=\beta(\tau)$. It immediately follows that, also with respect to the basis $\{ e_a\}$, the only nonzero independent components of the quadrupole tensor ($J_{abcd}$) are again $J_{0101}=J_{0202}$, $J_{0303}$, $J_{2323}=J_{1313}$, and $J_{1212}$, and that they are the same as those relative to $\{\eb_a\}$, {\it i.e.}, $J_{abcd}=\bar{J}_{abcd}$.
A frame of this kind which is well suited for doing calculations is constructed in the Appendix, which provides a practical way of obtaining the equations of motion by going to $\{ e_a\}$ first and then returning to the coordinate basis. 

As we also show in the Appendix, the components $\bar{J}_{abcd}$ of the quadrupole moment enter the equations of motion only via the combination
\begin{equation}\label{eq:qdef}
q:=\bar{J}_{0101}-\bar{J}_{0303}-\bar{J}_{2323}+\bar{J}_{1212}.
\end{equation}
The components $J_{0i0i}$ and $J_{ijij}$, $i=1,2,3$, are, respectively, the mass and stress quadrupole moments of the body~\cite{Ehlers1977DynamicsOE}. In this way, $q$ measures the imbalance of the mass or stress components of $T_{\mu\nu}$ between the $z$ direction and the $xy$ directions of the axisymmetric body under consideration. One might think of the body, then, as a prolate (oblate) ellipsoid when $q<0$ ($q>0$).%
\footnote{This is not strictly so since part of $q$ comes from the stress components of $J_{abcd}$, but it helps in making a mental image of the problem.}
For a spherically symmetric body we obviously have $q=0$ (this is also the case for a point particle, for which all the $J_{abcd}$ are zero).
This implies that a spinless, spherically symmetric test body follows geodesics in Schwarzschild spacetime, just as in the point-particle limit.

\subsection{Equations of motion}

Following the procedure outlined above we obtain the following equations of motion, valid up to quadrupole order (see the Appendix):
\begin{subequations}
\label{eq:sistdyn2}
\begin{align}
\frac{dr}{d\tau}      =&  f_1(r,p_r)+q(\tau)\,  g_1(r,p_r), \label{eq:sistdyn2a} \\
\frac{dp_r}{d\tau}  =& f_2(r,p_r)+q(\tau)\,  g_2(r,p_r), \label{eq:sistdyn2a}
\end{align}
\end{subequations}
in which $ f_1, f_2, g_1 $, and $g_2$ are given by \eqs{eq:fg}, where $E=-p_t$ and $L=p_\phi$.

A direct calculation shows that the following relations hold:
\begin{equation}\label{eq:constraintsfg}
\frac{\partial f_1}{\partial r}+\frac{\partial f_2}{\partial p_r}=0 \quad \text{and} \quad \frac{\partial g_1}{\partial r}+\frac{\partial g_2}{\partial p_r}=0,
\end{equation}
so that this dynamical system is Hamiltonian at each order, with canonical variables $(r,p_r)$. This allows us to use the simplest version of Melnikov's method to study the onset of chaos due to a periodic changing of the quadrupole moments of the body. Since Melnikov's method searches for chaos near unperturbed homoclinic orbits of the system, as explained below, we now look for them.

\subsection{\label{homorb}Point-particle case and homoclinic orbit} 
 
From the relationship $p_\mu p^\mu=-m^2$ we can write $p_r$ in terms of $r$ and $m$:
\begin{equation}
p_r^2=\frac{1}{f(r)} \left(
\frac{E^2}{f(r)}-\frac{L^2}{r^2}-m^2
\right),
\end{equation} 
where $m$ may (and usually does) vary with time when the body is extended. For a point particle, $m$ is constant, $m=m_0$, and the above equation can be used, in this case, to write $p_r$ as a function of $r$. Substituting this in \eqref{eq:sistdyn2a}, which in the point-particle case reads
$$
\frac{dr}{d\tau} =  f_1(r,p_r),
$$
we obtain
\begin{equation}
\label{eq:point_particle}
\frac{\dot{r}^2}{2}+\frac{1}{2}  \left(1-\frac{2 M}{r}\right) \left(\frac{l^2}{r^2}+1\right)=\frac{e^2}{2},
\end{equation}
where the specific energy and angular momentum, 
$$
e=E/m_0 \quad\text{and}\quad l=L/m_0,
$$
are also constants.%
\footnote{If the body is extended this is no longer true since in that case $m$ is not constant.}

\begin{figure}
	\begin{center}
		\includegraphics[width=0.75\columnwidth]{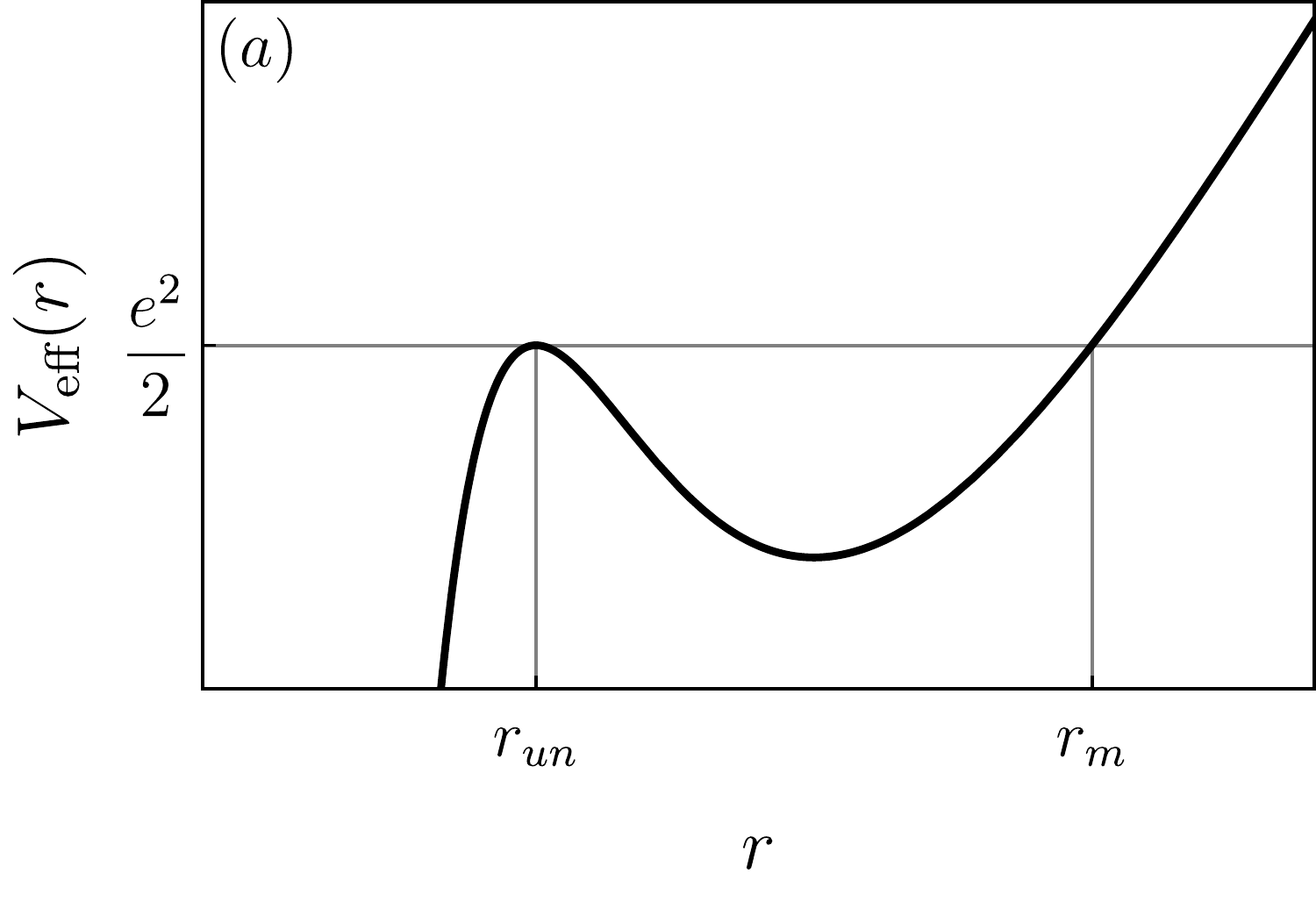}
		\includegraphics[width=0.75\columnwidth]{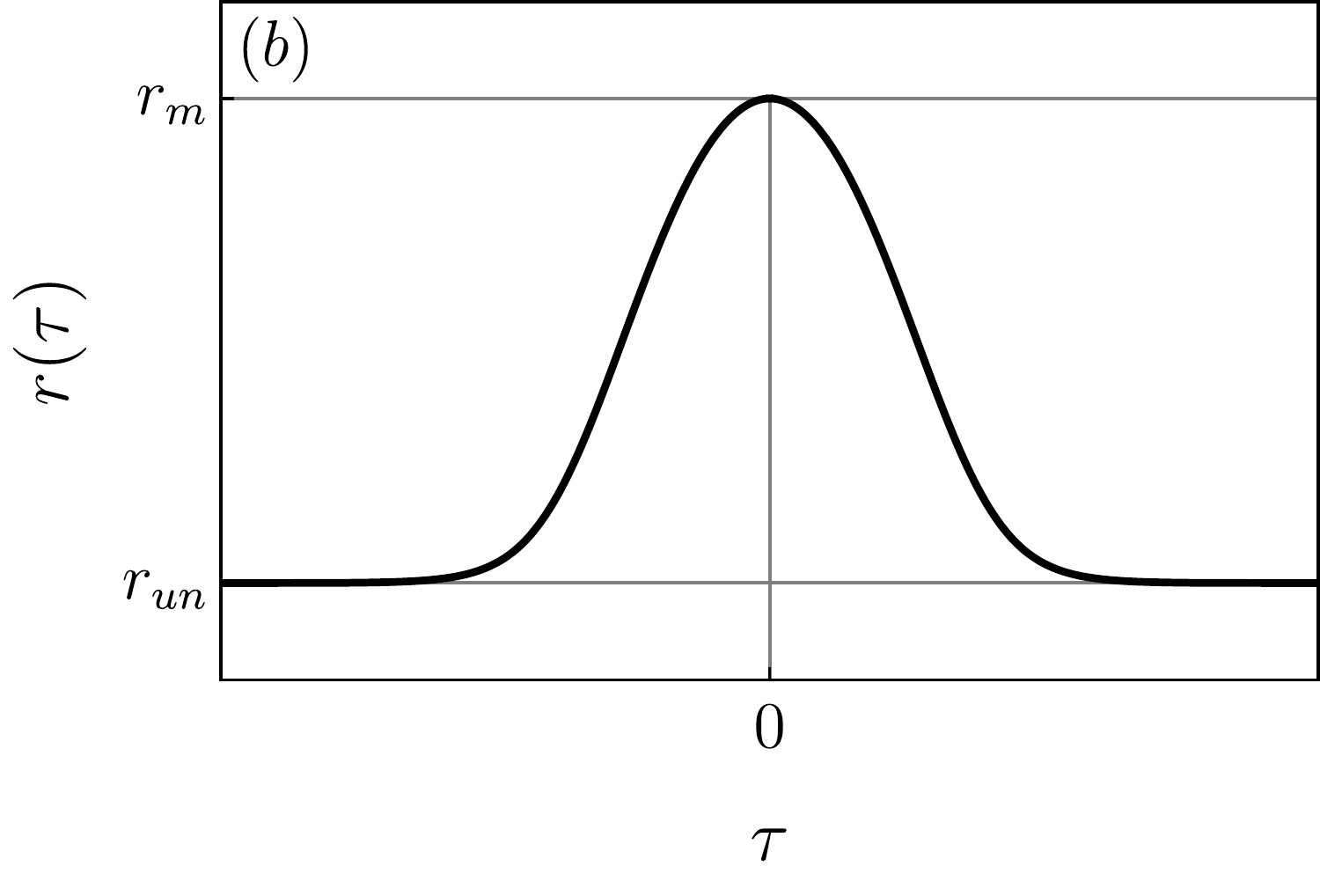}  		
		\caption{
		(a) Effective potential for the point-particle case. The orbit with ``energy'' $e^2/2=V_{\rm eff} (r_{un})$ is a homoclinic orbit. 
		(b) Plot of $r(\tau)$ for a homoclinic orbit.
		}
		\label{fig:orbhom}
	\end{center}
\end{figure}

Equation (\ref{eq:point_particle}) is the usual conservation equation for the orbits of massive particles in Schwarzschild spacetime when written in terms of an ``effective potential'' 
$$
V_{\text{eff}}=\frac{1}{2}-\frac{M}{r}+\frac{l^2}{2 r^2}-\frac{M l^2}{r^3}
$$
and with ``energy'' $e^2/2$. For a certain range of values of energy and angular momentum, $V_{\text{eff}}$ has a local maximum which corresponds to an unstable circular orbit at a certain $r=r_{un}$ with $3M < r_{un}< 6M$ (see Fig.~\ref{fig:orbhom}(a)). 
If we consider only bounded energy level sets ($e<1$), then we must restrict this range to $4M < r_{un}< 6M$.

The radius $r_{un}$ is determined by the (constant) angular momentum $l$. This relation can be inverted to parametrize the unstable circular orbit by $r_{un}$, whose energy and angular momentum are given by 
$$
e=\frac{r_{un}-2 M}{\sqrt{r_{un} \left(r_{un}-3 M\right)}}
\quad \text{and} \quad
 l = \frac{\sqrt{M} r_{un}}{\sqrt{r_{un}-3 M}}. 
$$ 
In the phase space $(r, p_r)$, the curve that starts and ends at $ r_{un} $ corresponds to the homoclinic orbit that we want to determine. The turning point for this orbit occurs at
$$ r_{m} =\frac{2 M r_{un}}{r_{un}-4 M}.$$ 
We note that the turning point radius $r_m$ is larger the closer $r_{un}$ is to $4M$.
Equation (\ref{eq:point_particle}) yields
\begin{equation}\label{eqmovcaos}
    \frac{r^{3/2}\dot{r}}{(r-r_{un})\sqrt{r_m-r}}=\pm \nu,
\end{equation}
with 
$$
\nu = \sqrt\frac{M (r_{un} -4M)}{r_{un} (r_{un} -3M)}.
$$
This leads to
\begin{equation}\label{orbhom}
\begin{split}
 \pm \nu \tau =& \sqrt{r(r_m-r)}+(r_m+2 r_{un})\arctan \sqrt{\frac{r_m-r}{r}}\\
 &+\frac{2 r_{un}^{3/2}}{\sqrt{r_m-r_{un}}}\arctanh\sqrt{\frac{r_{un}(r_m-r)}{r(r_m-r_{un})}},
 \end{split}
\end{equation}
in which the integration constant was chosen in such a way that $ \tau (r = r_m) = 0 $. Figure~\ref{fig:orbhom}(b) shows a typical plot of $ r $ versus $ \tau $.
We note that $ r(\tau) $ tends asymptotically to the unstable equilibrium point $ r_{un} $.

\subsection{Extended body and quadrupolar oscillations}

We now go back to the case of an extended body which is axisymmetric and has reflection symmetry with respect to the equatorial plane. We saw that in the quadrupolar approximation the finite-size corrections enter the equations of motion (\ref{eq:sistdyn2}) through the parameter $q=q(\tau)$ of \eq{eq:qdef}.

In order to analyze the effects of a changing-shape configuration of the test body on its own translational dynamics, Eqs.~(\ref{eq:sistdyn2}), we consider a time-dependent $q$ with frequency $\Omega$, 
\begin{equation}\label{eq:qcos}
q(\tau) = q_0 \sin\left(\Omega\tau\right),
\end{equation}
oscillating between an oblate ($q>0$) and a prolate ($q<0$) spheroid. We show next that this leads to homoclinic chaos for the extended body if the nonperturbed orbit (in the point-particle limit) is taken as a homoclinic orbit.

\section{\label{melnikov} Homoclinic intersections due to finite-size effects }

We start by briefly reviewing Melnikov's method, an analytical tool that allows us to search for homoclinic intersections in the perturbed system's phase space. Consider a two-dimensional system subjected to a time periodic perturbation, for which the equations of motion can be written in the form 
\begin{equation}\label{eq:EDOMelnikov}
\dot{\bm{x}}=\bm{f}(\bm{x})+\epsilon \bm{\lambda}(\bm{x},\tau),
\end{equation}
where $\bm{x}=(r,p_r)$, $\bm{f}=(f_1(r,p_r),f_2(r,p_r))$, $\bm{\lambda}=(\lambda_1(r,p_r,\tau),\lambda_2(r,p_r,\tau))$, a dot represents differentiation with respect to $\tau$, and $\epsilon \ll 1$. Assume that $\bm{f}$ and $\bm{\lambda}$ are smooth and bounded and that $\boldsymbol{\lambda}$ is periodic in $\tau$. Also assume that the unperturbed system ($\epsilon=0$) is Hamiltonian (therefore integrable) and that $(r, p_r)$ are canonical variables. Then
\begin{equation}\label{eq:constraintfMelnikov}
    \frac{\partial f_1}{\partial r}+\frac{\partial f_2}{\partial p_r}=0. 
\end{equation}
Now suppose that the unperturbed system has an unstable equilibrium point $P_0$ and an associated homoclinic orbit in phase space. Since the system is integrable, the system's phase portrait will present continuous curves in the $(r,p_r)$ space. 

Also, consider that the perturbation is Hamiltonian. Then we can construct a two-dimensional stroboscopic Poincar\'e map by plotting in the $(r,p_r)$ plane, for each orbit, the events $t_n=2n\pi/\Omega$ associated with a fixed numerical value of the Hamiltonian (so that the map is symplectic). For sufficiently small $\epsilon$, the fixed point in the map is displaced but still exists. The homoclinic orbit generally splits into a stable and an unstable manifold of the fixed point according to the map; if they present transverse intersections in the $(r,p_r)$ plane, then the integrability of the system is broken \cite{lichtenbergLieberman1992, holmes}.

One way to search for these homoclinic intersections between the stable and unstable manifolds is by means of Melnikov's integral, which gives us the first-order term in the transverse distance (with respect to the $\tau=\tau_0$ point of the unperturbed homoclinic orbit), in phase space, between these two sets. 
A complete derivation of Melnikov's integral can be found in Ref. \cite{Holmesbook}. For our purpose, it is important to know that
since Eq.~(\ref{eq:constraintfMelnikov}) holds, Melnikov's integral is given by \cite{holmes}
\begin{equation}\label{melnikov2}
    M(\tau_0)=\int_{-\infty}^{\infty}(f_1\lambda_2-f_2\lambda_1)(r(\tau),p_r(\tau),\tau+\tau_0)\,d\tau,
\end{equation}
in which $(r(\tau),p_r(\tau))$ is evaluated along the unperturbed homoclinic orbit. Therefore if $M(\tau_0)$ has isolated zeros then the unstable and stable manifolds of the perturbed fixed point cross transversely, generating a homoclinic tangle in phase space \cite{lichtenbergLieberman1992}.

In our case, $f_1$ and $f_2$  are given by Eqs.~(\ref{eq:f1A}) and (\ref{eq:f2A}), and $\epsilon\lambda_1=q(\tau)\,g_1$, $\epsilon\lambda_2=q(\tau)\,g_2$, with $q(\tau)$ given by Eq.~(\ref{eq:qcos}) and $g_1,g_2$ given by Eqs.~(\ref{eq:g1A}) and (\ref{eq:g2A}). Here, $q_0$ plays the role of the small perturbation parameter $\epsilon$ in Eq.~(\ref{eq:EDOMelnikov}).
Then the system is Hamiltonian [\eqs{eq:constraintsfg} are satisfied] and Melnikov's method can be applied to the system, whose unperturbed homoclinic orbit is given by Eq.~(\ref{orbhom}). After a lengthy calculation, the Melnikov integral (\ref{melnikov2}) can be written as
\begin{equation}\label{eqk}
\begin{split}
 M(\tau_0)=2 \cos{(\Omega \tau_0)}K(\Omega) 
    \end{split},
\end{equation}
where
\begin{equation}
K(\Omega)=\int_{r_{un}}^{r_m}k(r)\sin{[\Omega \, \tau(r)]}dr,
\end{equation}
with 
\begin{equation}
k(r)=\frac{2 M \left(-3 M r^2+5 M r_{un}^2+r^2 r_{un}\right)}{r^6 \left(r_{un}-3 M\right)}\,.
\end{equation}

\begin{figure}[!h] 
    \includegraphics[width=\columnwidth]{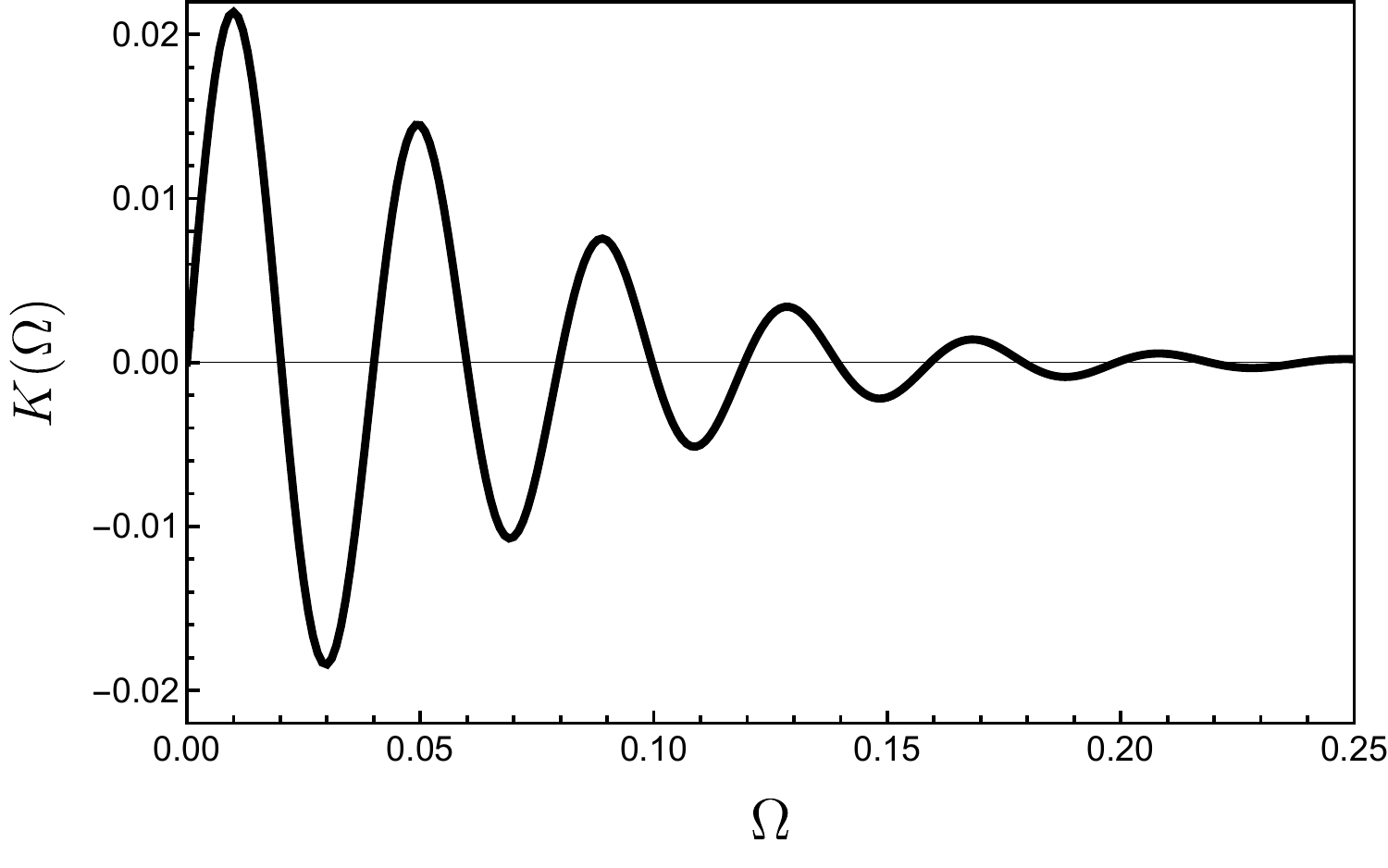}
    \includegraphics[width=\columnwidth]{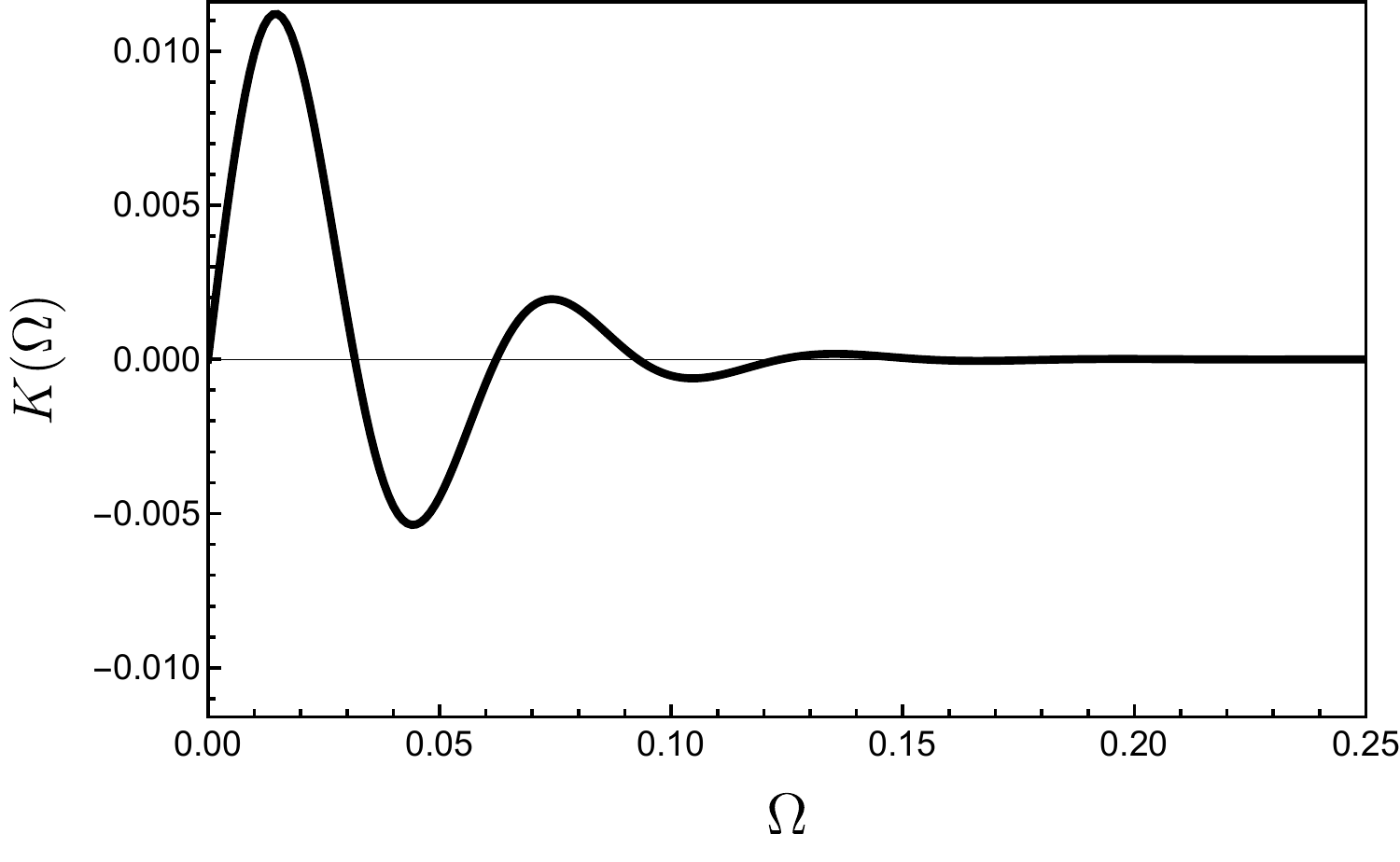}
\caption{The function $K(\Omega)$ for $r_{un}=4.5 M$ (top) and $r_{un}=5 M$ (bottom); units are such that $M=1$. We see that $K(\Omega)\ne0$ almost everywhere in both plots.}
    \label{grafK}
\end{figure}

If $\Omega$ is such that $ K(\Omega)\ne 0 $, the Melnikov integral will have only isolated simple roots, which will be located at $ \Omega \tau_0 = (n-1/2) \pi $, with $n$ integer.
The behavior of $ K $ as a function of $ \Omega $ for two different values of $ r_{un} $ can be seen in Fig.~\ref{grafK}, which shows that $K(\Omega)= 0$ solely at a discrete set of values of $\Omega$. This implies the existence of a homoclinic tangle for arbitrarily small values of the perturbation $q_0$. We note that both the amplitude and the number of oscillations  increase as $ r_{un} $ decreases, {\it i.e.}, as the unperturbed unstable circular orbit gets closer to the black hole.

\begin{figure}[!h] 
    \includegraphics[width=.8\columnwidth]{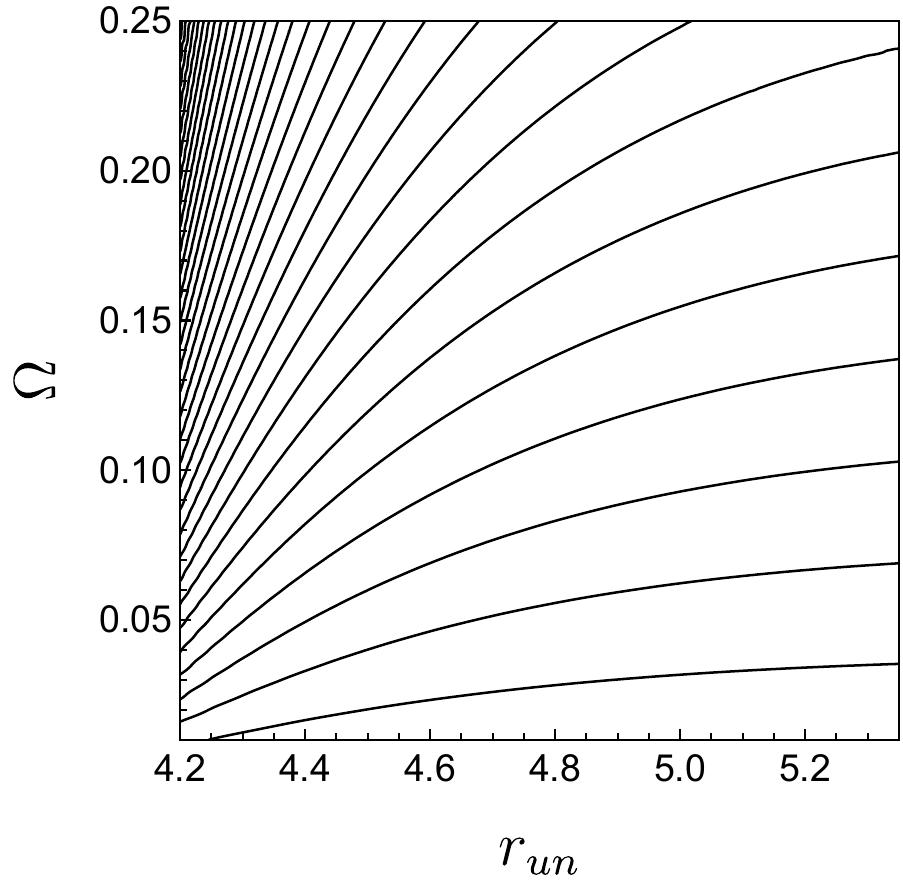}
\caption{Contour curves corresponding to $K=0$ in the $(r_{un},\Omega)$ plane; units are such that $M=1$. We see that $K\ne0$ almost everywhere in this domain.}
    \label{curvani0}
\end{figure}

The behavior of $ K $ can also be illustrated by its $K=0$ contour plot in the $(r_{un},\Omega)$ plane, Fig.~\ref{curvani0}. This shows the generic appearance of homoclinic intersections, no matter how small the quadrupole perturbation to the point-particle dynamics is.

A question which arises is whether these homoclinic intersections generate chaotic behavior for infinite proper time. For Hamiltonian systems with bounded unperturbed energy level sets, homoclinic chaos is guaranteed once homoclinic intersections are detected \cite{lichtenbergLieberman1992,holmes,Holmesbook}. In our case, on the other hand, the Hamiltonian system can be seen as a system with an escape (or exit, see Ref.~\cite{bleherEtal1998PRA}), the escape being the infall of the body into the black hole. Therefore the system exhibits transient chaos \cite{ott2002chaos, dasilvaEtal2002PhPl}, 
since trajectories will either be regular and bounded or will eventually fall into the black hole (except from sets of measure zero, which do not fall into the black hole, as for instance those with initial conditions lying at the homoclinic tangle associated with the unstable fixed point \cite{ott2002chaos, dasilvaEtal2002PhPl, demouraLetelierPRE2000}). 

It is implicit in our assumptions about the controllability of the body's internal distribution that, no matter how strong the tidal forces acting on it are, the internal mechanisms will be able to balance them and maintain the prescribed oscillations. Also, higher order multipole terms were neglected in our analysis. These considerations would have to be taken into account in a more realistic treatment of the problem.

\section{Conclusion}
\label{sec:conclusion}

It has been recently shown that periodic oscillations in the mass distribution of a nonrelativistic test body may generate homoclinic chaos in its otherwise integrable translational dynamics~\cite{bolinha}. Here we took advantage of the homoclinic orbit associated with an unstable circular orbit in Schwarzschild spacetime to show that a similar effect appears in general relativity (notice that such orbits have no analog in the corresponding nonrelativistic Kepler's problem).

Specifically, we considered a spinless axisymmetric spheroid aligned with the equatorial plane of the spacetime and allowed it to change its shape periodically between oblate and prolate configurations. We then showed that this breaks the system's integrability via homoclinic crossings of the stable and unstable manifolds associated with the unstable circular orbit. These crossings are identified via Melnikov's method and give rise to transient chaos in the system. All of this was done at quadrupolar order using Dixon's formalism for extended bodies in general relativity.

It is interesting to note that several authors studied homoclinic chaos for test-particle motion on black-hole spacetimes in the past decades~(see, for instance, Refs.~\cite{bombelli,letelierrel}). Our approach is conceptually different from theirs as they consider point particles subjected to a spacetime with an externally perturbed metric.  In the present work, the spacetime is the pure, unperturbed Schwarzschild spacetime, and the perturbation to test-particle motion comes from the finite-size structure of the test body. 

We therefore believe that the present approach opens new windows to the study of the dynamics of extended bodies in standard spacetimes of general relativity.
The present model is applicable to any object with time-varying internal structure, once the quadrupole approximation for the body is valid. Although the perturbation assumed in this work is sinusoidal, any time-dependent quadrupole perturbation $q(\tau)$ is encompassed by the formalism. This is the case, for instance, of the long-term behavior of astronomical bodies with varying internal structure, such as the orbital motion of variable stars with nonradial pulsating patterns and satellite motion with prescribed (time-dependent) shape-changing variations.

Regarding the orbital motion of pulsating stars, although the typical period of revolution of a star in the Milky Way is of the order of 100 Myr, this period becomes very short near a supermassive black hole as the object approaches the light ring, $r=3M$. Therefore it is possible, in principle, to have imprints of the chaotic nature of the orbital motion of such bodies in the time series of the brightness variability of these stars.

\begin{acknowledgments}
The authors acknowledge insightful discussions with R. D. Vilela. R.A.M. was partially supported by Conselho Nacional de Desenvolvimento Cient\'{i}fico e Tecnol\'{o}gico under Grant 310403/2019-7.  F.F.R. acknowledges support from Coordena\c{c}\~{a}o de Aperfei\c{c}oamento de Pessoal de N\'{i}vel Superior (CAPES, Brazil) Grant No. 88882.329025/2019-01.
\end{acknowledgments}

\appendix

\section{Obtaining the equations of motion}

Here we construct a moving frame $\{ e_a\}$ along the center-of-mass worldline $z^\mu(\tau)$ which is more amenable to calculations than the M-transported frame, $\{\eb_a\}$, but still has the property $J_{abcd}=\bar{J}_{abcd}$ (see main text). 

We first take $e_0=u$ and $e_3=-e_{\hat{\theta}}$ and choose $e_1$ and $e_2$ as follows. 
Let $\hat{\nu}_u$ be the unit vector along the projection of $u$ on the rest space of $e_{\hat{t}}$, so that $\hat{\nu}_u=\cos(\alpha_u) e_{\hat{r}}+\sin(\alpha_u) e_{\hat{\phi}}$, for some angle $\alpha_u$. Then $u=\gamma_u \left( e_{\hat{t}} +\nu_u \hat{\nu}_u  \right)$, with $\nu_u$ a three-(dynamical velocity) and $\gamma_u=(1-\nu_u^2)^{-1/2}$. We may then define $e_1= \sin (\alpha_u)  e_{\hat{r}}- \cos (\alpha_u) e_{\hat{\phi}}$ and $e_2= \gamma_u \left( \nu_u e_{\hat{t}}+\hat{\nu}_u \right)$. The M-transported frame $\{\eb_{a}\}$ must then differ from $\{ e_a\}$ only by a rotation in the plane of $e_1$ and $e_2$, so that $\eb_1=\cos(\beta) e_1+\sin(\beta) e_2$, $\eb_2=-\sin(\beta) e_1+\cos(\beta) e_2$, with $\beta=\beta(\tau)$. 
It immediately follows, also in the basis $\{ e_a\}$, that the only nonzero independent components of the quadrupole tensor ($J_{\alpha\beta\gamma\delta}$) are  $J_{0101}=J_{0202}$, $J_{0303}$, $J_{2323}=J_{1313}$ and $J_{1212}$, and, for all of these, we get $J_{abcd}=\bar{J}_{abcd}$.%
\footnote{
It is worth noting that the quadrupole moments also depend on the choice of the evolution parameter $s$. If $J_{\alpha\beta\gamma\delta}$ ($\tilde{J}_{\alpha\beta\gamma\delta}$) is given as if the evolution parameter $s$ was chosen as $\tau$ ($\tilde{\tau}$), we have that $J(\tau)= \frac{d\tilde{\tau}}{d\tau} \tilde{J}(\tilde{\tau})$. The factor in front of $\tilde{J}_{\alpha\beta\gamma\delta}$ is anyway unimportant if one is interested in working up to octupole order, since in that case  $\frac{d\tilde{\tau}}{d\tau} \tilde{J}=v_\mu u^\mu \tilde{J}$ may be simply replaced by $\tilde{J}$~\cite{monografiaRodrigo}. 
}

One can write $\nu_u$, $\alpha_u$ and $m$ in terms of $p_t$, $p_r$ and $p_\phi$ as follows. By definition of the frame $\{e_a\}$, we have that $(p_a)=(-m,0,0,0)$. On the other hand, $p_a=(e_a)^\mu p_\mu$. From this equality between the two forms of writing $p_a$ we get
\begin{subequations}
\label{eq:}
\begin{align}
\nu_u \cos(\alpha_u)=&  -\frac{p_r}{p_t} f(r), \\
\nu_u \sin(\alpha_u)=&  -\frac{p_\phi}{p_t} \frac{\sqrt{f(r)}}{r},
\end{align}
\end{subequations}
along with the already known relation $m^2=-p_\mu p^\mu=f(r)^{-1} p_t^2-f(r) p_r^2 - r^{-2} p_\phi^2$. Calculating the force in \eq{eq:psimplificada} in the moving frame (wherein $J$ is simplest) and transforming it back to the coordinate basis we obtain the following nonzero components of $F_\mu$:
\begin{subequations}
\label{eq:}
\begin{align}
F_r =&
\frac{2 M q}{r^4}
\left[
3 p_\phi^2 \left(1-\tfrac{2 M}{r}\right) A-1
\right], \\
F_\phi =&
\frac{4 M q}{r^2}
p_r p_\phi \left(1-\frac{2 M}{r}\right)^2 A,
\end{align}
\end{subequations}
with $A=\tfrac{1}{\left(1-\frac{2 M}{r}\right) \left(p_r^2 r^2 \left(1-\frac{2 M}{r}\right)+p_\phi^2\right)-p_t^2 r^2}$. Substituting this in \eq{eq:psimplificada} allows us to find the right-hand sides of \eqs{eq:U=something}:
\begin{widetext}
\begin{subequations}
\label{eq:}
\begin{align}
\frac{dt}{d\tau} =&
\frac{E (r-2 M) \left(L^2 (r-2 M) \left(8 M q-r^4 \dot{p}_r\right)+r \left(2 M q+r^4 \dot{p}_r\right) \left(E^2 r^2-(r-2 M)^2 p_r^2\right)\right)}{(r-2 M)^3 \left(M r p_r^2-L^2\right) \left(L^2+r (r-2 M) p_r^2\right)-E^2 L^2 r^3 (M-r) (r-2 M)-E^4 M r^6}, \\
\frac{dr}{d\tau} =&
\frac{(r-2 M)^3 p_r \left(L^2 (r-2 M) \left(8 M q-r^4 \dot{p}_r\right)-r \left(2 M q+r^4 \dot{p}_r\right) \left((r-2 M)^2 p_r^2-E^2 r^2\right)\right)}{r^2 (r-2 M)^3 \left(M r p_r^2-L^2\right) \left(L^2+r (r-2 M) p_r^2\right)+E^2 L^2 r^5 (r-M) (r-2 M)-E^4 M r^8}
,\\
\frac{d\phi}{d\tau} =&
\frac{L (r-2 M) \left((r-2 M)^2 \left(-\left(r^4 \dot{p}_r \left(L^2+r (r-2 M) p_r^2\right)\right)+4 L^2 M q+2 M q r (4 M-r) p_r^2\right)+E^2 r^4 \left(r^3 (r-2 M) \dot{p}_r+2 M q\right)\right)}{r^3 (r-2 M)^3 \left(L^2+r (r-2 M) p_r^2\right) \left(M r p_r^2-L^2\right)+E^2 L^2 r^6 (r-M) (r-2 M)-E^4 M r^9}
,
\end{align}
\end{subequations}
\end{widetext}
where $E=-p_t$ and $L=p_\phi$.
Noticed that none of these expressions have $t$ or $\phi$ on their right-hand side, as anticipated in the main text. They do depend on $\tau$ though, via $q=q(\tau)$. 

Finally, following the procedure of the main text we obtain, up to quadrupole order, the dynamical system \eqref{eq:sistdyn2} with
\begin{widetext}
\begin{subequations}
\label{eq:fg}
\begin{align}
f_1 =&
\frac{(r-2 M)^{3/2} p_r}{\sqrt{E^2 r^3-(r-2 M) \left(L^2+r (r-2 M) p_r^2\right)}}, \label{eq:f1A} \\
g_1= & 
-\frac{4 L^2 M (r-2 M)^3 p_r}{\left(r (r-2 M) \left(L^2+r (r-2 M) p_r^2\right)-E^2 r^4\right){}^2} \label{eq:g1A}
, \\
f_2 =&
\frac{(r-2 M)^2 \left(L^2-M r p_r^2\right)-E^2 M r^3}{r^2 (r-2 M)^{3/2} \sqrt{E^2 r^3 - (r-2 M) \left(L^2+r (r-2 M) p_r^2\right)}}
, \label{eq:f2A}\\
g_2= & 
\frac{2 M \left(2 L^4 (r-2 M)^2+L^2 r \left((r-2 M)^2 (3 r-4 M) p_r^2+E^2 r^2 (8 M-3 r)\right)-\left(r (r-2 M)^2 p_r^2-E^2 r^3\right){}^2\right)}{r^4 \left((r-2 M) \left(L^2+r (r-2 M) p_r^2\right)-E^2 r^3\right){}^2}
. \label{eq:g2A}
\end{align}
\end{subequations}
\end{widetext}


\providecommand{\noopsort}[1]{}\providecommand{\singleletter}[1]{#1}%
%


\end{document}